\documentclass[amsmath,amssymb,aps,prl,twocolumn,showkeys,superscriptaddress,10pt,floatfix,showpacs]{revtex4-1}

\usepackage{tabularx}
\usepackage[dvipsnames]{xcolor}
\usepackage{fdsymbol}
\usepackage{graphicx}
\usepackage{bm}

\newcommand{\TableEntry}[3]{
	\begin{minipage}[t][0.3in]{0.4in}
		#1\newline (${\color{#2} #3}$)
	\end{minipage}
}

\begin{document}

\title{Superconductivity of barium-VI synthesized via compression at low temperatures}

\author{D.\ E.\ Jackson} 
\affiliation{Department of Physics, University of Florida, Gainesville, FL 32611}
\author{D.\ VanGennep}
\affiliation{Department of Physics, University of Florida, Gainesville, FL 32611}
\author{Y.\ K.\ Vohra}
\affiliation{Department of Physics, University of Alabama at Birmingham, Birmingham, AL, 35294}
\author{S.\ T.\ Weir}
\affiliation{Physics Division, Lawrence Livermore National Laboratory, Livermore, CA 94550, USA}
\author{J.\ J.\ Hamlin}
\email{Corresponding author: jhamlin@ufl.edu}
\affiliation{Department of Physics, University of Florida, Gainesville, FL 32611}

\date{\today}

\begin{abstract}
Using a membrane-driven diamond anvil cell and both ac magnetic susceptibility and electrical resistivity measurements, we have characterized the superconducting phase diagram of elemental barium to pressures as high as 65~GPa.  We have determined the superconducting properties of the recently discovered Ba-VI crystal structure, which can only be accessed via the application of pressure at low temperature.  We find that Ba-VI exhibits a maximum $T_c$ near 8~K, which is substantially higher than the maximum $T_c$ found when pressure is applied at room temperature.\end{abstract}

\maketitle

\section{Introduction}
Experiments under applied high pressures have played an important role in advancing our understanding of relationships between crystal structure and superconducting properties~\cite{Schilling2001,Hamlin2015}.  It is well known that the critical pressures for structural transitions are sometimes  strongly temperature dependent~\cite{Cannon1974} and therefore one must use caution when comparing room temperature structural information to trends in (usually low temperature) superconducting critical temperatures~\cite{Matsuoka2014}.  Furthermore, in some cases, the thermodynamic history of the sample can strongly influence superconducting~\cite{Veal1997} and/or structural~\cite{Deemyad2017} properties.  For example, if there is a kinetic barrier for the transition to the stable phase, applying or releasing pressure at low temperatures can lead to a crystal structure persisting far outside of the range of pressures for which it is thermodynamically stable~\cite{Gregoryanz2001}.  An arguably more dramatic manifestation of such effects can occur if a barrier for the transition to the stable phase causes an entirely new, metastable crystal structure to emerge.  This was recently found for elemental Ba metal, with the observation of a new, metastable structural phase (Ba-VI) that only appears when pressure is applied at temperatures below $\sim 100$~K~\cite{Desgreniers2015}.  In this work we have determined the superconducting properties of this newly discovered phase of Ba.  The superconducting properties of the alkaline earth elements (Be, Mg, Ca, Sr, Ba, Ra) are of particular interest because Ca at high pressure exhibits the highest superconducting critical temperature of any of the elemental solids with $T_c \sim 21-29$~K at pressures above 200~GPa~\cite{Okada1996,Yabuuchi2006,Sakata2011,Andersson2011}.

The room temperature crystal structures of Ba at high pressure have been studied to pressures as high as 110~GPa~\cite{Kenichi1994,Winzenick1996,Winzenick1997}.  These studies revealed structural phase transitions from Ba-I (bcc) at ambient pressure to Ba-II (hcp), Ba-IV, and Ba-V (hcp) at pressures of 5~GPa, 12 GPa, and 45~GPa, respectively.  Early reports of a Ba-``III'' phase, existing between Ba-II and Ba-IV, appear to have been refuted, though a new phase, also labeled Ba-III may form in a very narrow temperature and pressure range ($\sim 5-7$~GPa at $\sim 700-800$~K)~\cite{Winzenick1997}.  Ba-IV in particular has received a great deal of attention because of its extremely complex modulated ``host-guest'' structures, comprised of coexisting inter-penetrating, sub-lattices~\cite{Loa2012}.  The label ``Ba-IV'' actually refers to a series of closely related structures (Ba-IVa through Ba-IVd) built upon the same host-guest structural motif.  It is within the pressure range that Ba-IV exists that Ba exhibits a maximum $T_c$ value of $\sim 5$~K at 18 GPa.

\begin{figure}
\includegraphics[width=\columnwidth]{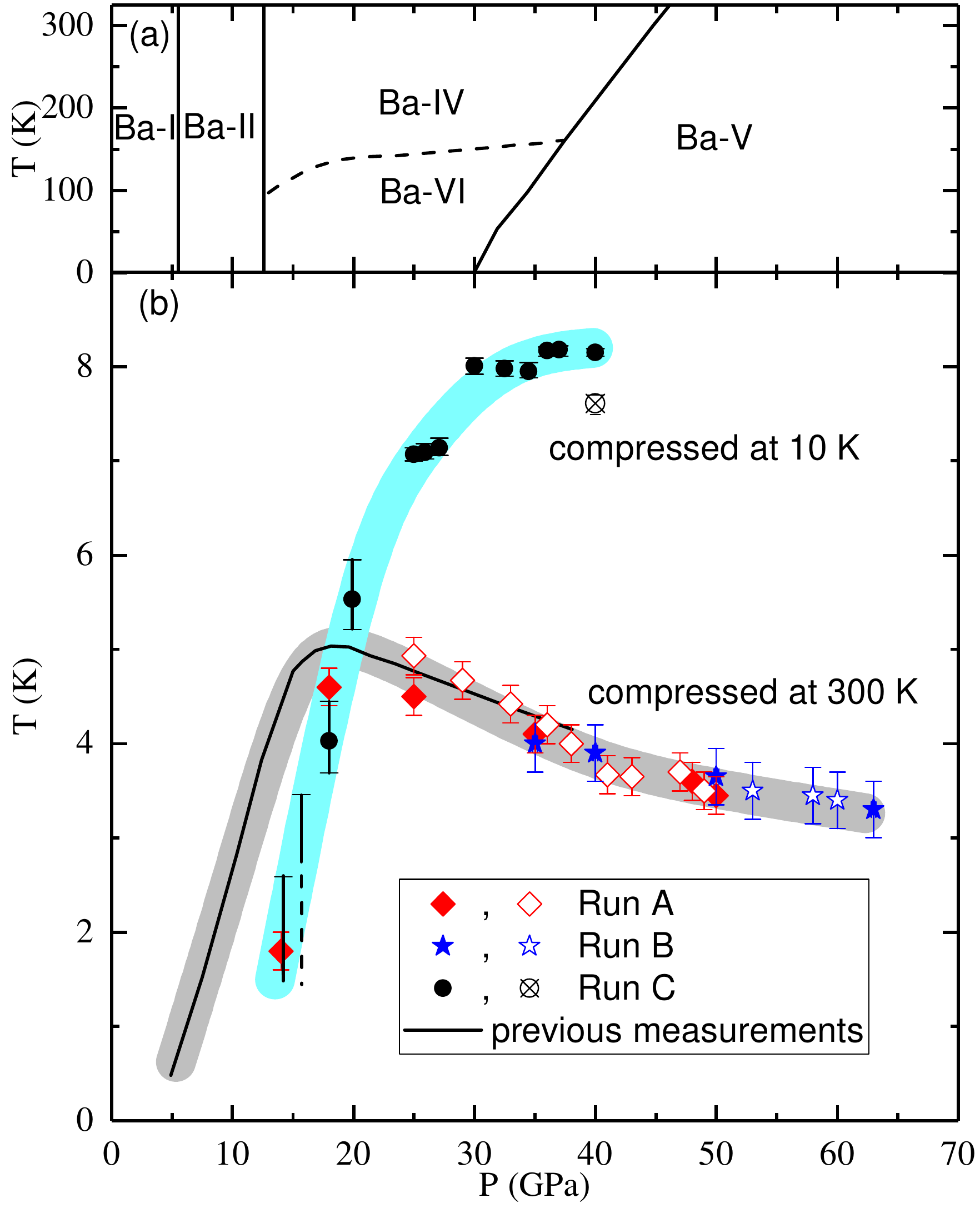}
\caption{(a) Schematic structural phase diagram for Ba based on Ref.~\cite{Desgreniers2015}. (b) Phase diagram indicating superconducting critical temperature versus pressure.  The solid black line indicates $T_c$ as determined by previous measurements in which pressure was varied at room temperature~\cite{Wittig1969,Moodenbaugh1973,Dunn1982}.  The $T_c$ of Ba-VI (black circles) reaches substantially higher values than any of the other structures of Ba.  Further details of each measurement run are summarized in Table~\ref{table:run_summary}.}
\label{fig:phase}
\end{figure}

When Ba is compressed at temperatures below $\sim 100$~K, a substantial energy barrier inhibits the Ba(II$\rightarrow$IV) transition and instead a new structure, denoted Ba-VI, appears~\cite{Desgreniers2015}.  Unlike complex Ba-IV, Ba-VI adopts a rather simple orthorhombic structure.  The Ba-VI structure exists for pressures in the range of 12-30~GPa at low temperature, and the structure appears to be metastable, as evinced by the fact that on warming above $\sim 100$~K it transforms to Ba-IV.  Figure~\ref{fig:phase}a schematically represents the structural phase diagram reported in Ref.~\cite{Desgreniers2015}.

We emphasize that Desgreniers~\textit{et al}.~\cite{Desgreniers2015} found that Ba-VI cannot be synthesized by cooling from Ba-IV.  Previous studies of the superconducting properties of Ba have nearly all been carried out by applying pressure at room temperature, and then cooling at constant pressure, so that the most widely reported and quoted $T_c$ values correspond to Ba-IV rather than Ba-VI~\cite{Wittig1969,Moodenbaugh1973,Dunn1982}.  There are two exceptions.  In the 1972 report of Il'ina~\textit{et al}.~\cite{Ilina1972} pressures up to 19 GPa were applied at 4.2~K, resulting in critical temperatures about 0.5~K lower than found for room temperature pressure application.  In 1989, Bireckoven and Wittig~\cite{Bireckoven1989,Tonkov1996} described an experiment in which they released pressure from 50~GPa at low temperatures.  They found that below 35~GPa, $T_c$ began to increase more rapidly, eventually reaching up to 7~K at 27~GPa.  They attributed the higher critical temperature to the metastability of Ba-V on pressure release.  However, with the new structural information~\cite{Desgreniers2015} one may wonder if perhaps the increase in $T_c$ was due to the formation of Ba-VI.

In the present study, we have performed measurements up to $\sim 65$~GPa using a membrane driven diamond anvil cell which allows pressure changes at low temperature.  In order to unmask the superconducting properties of Ba-VI we have used a combination of ac magnetic susceptibility (ACS) measurements, which are sensitive to the superconductivity of the bulk phase, and electrical resistivity measurements, which are sensitive the superconductivity of small amounts of minority phase.  We find that Ba-VI exhibits a maximum $T_c$ near 8~K, which is significantly higher that the highest value reported for Ba-IV (5~K).  We have also performed ACS measurements of the pressure dependence of $T_c$ for Ba-V in the vicinity of the Ba(V$\rightarrow$VI) transition by unloading pressure at low temperature.  We find no evidence for the substantial $T_c$ enhancement in metastable Ba-V suggested by Bireckoven and Wittig~\cite{Bireckoven1989,Tonkov1996}.  Together with Ba, a number of other elements exhibit complex crystal structures with commensurate or incommensurate modulations at high pressure~\cite{McMahon2006} and we discuss what our results might imply for the superconducting phase diagrams of such systems.

\section{Methods}
\begin{table}
\begin{tabularx}{\linewidth}{|l|l|X|}
\hline
\textbf{Run} & \textbf{Type} & \textbf{Measurement sequence} \\
\hline
\TableEntry{Run A}{red}{\medblackdiamond,\meddiamond} & $\chi$ & Measurements up to $\sim 49$~GPa performed while applying pressure at room temperature (filled diamond).  Unloading measurements performed while maintaining $T\lesssim 10$~K (open diamond).\\
\hline
\TableEntry{Run B}{blue}{\medblackstar,\medwhitestar} & $\chi$ & Measurements up to $\sim 63$~GPa performed while applying pressure at room temperature (filled star).  Unloading measurements performed while maintaining $T\lesssim 10$~K (open star).\\
\hline
\TableEntry{Run C}{black}{\medblackcircle,\otimes} & $\rho$ & Pressure increased to 8~GPa at room temperature and then cooled.  Further measurements performed during pressure application below 10~K (filled circle).  At the highest pressure, the sample was ``annealed'' at 50~K for three hours before re-measuring at the same pressure (circle with cross).\\
\hline
\end{tabularx}
\caption{Summary of the different sets of measurements. The symbols in the leftmost column correspond to the symbols used in the phase diagram presented in Fig.~\ref{fig:phase}. The column labeled ``Type'' indicates whether resistivity ($\rho$) or AC magnetic susceptibility ($\chi$) was measured.  ``Measurement sequence'' indicates the path that was taken through temperature-pressure phase space.}
\label{table:run_summary}
\end{table}
Pieces of Ba with typical dimensions around $100\, \mu \mathrm{m} \times 100\, \mu \mathrm{m} \times 10\, \mu \mathrm{m}$ were cut from a large piece of 5N purity Ba. The sample was cut and loaded into the diamond anvil cell inside of a nitrogen-filled purified glovebox.  The measurements were made in an OmniDAC gas membrane-driven diamond anvil cell from Almax-EasyLab. The cell was placed inside a custom, continuous flow cryostat built by Oxford instruments. A fiber entering the top of the cryostat via a feed-through and windows on the bottom of the cryostat provide optical access to the cell. The pressure was determined using either the fluorescence of the \textit{R}$_1$ peak of small ruby chips placed next to the sample~\cite{Piermarini1975} or from the Raman signal from the anvil~\cite{Hanfland1985}.  The gas membrane and optical access allows us to change and measure pressure \textit{in situ} at low temperatures.

Resistivity was measured by using a designer diamond anvil with eight symmetrically positioned, deposited tungsten microprobes encapsulated in high quality homo-epitaxial diamond~\cite{Weir2000}. This diamond had a culet of approximately 180 $\mu$m. Solid steatite was used as pressure medium. Resistance was measured in a four probe arrangement using Keithley 6221 and Keithley 2182a configured for ``delta mode.'' From the resistance, $R$ we estimated the sample resistivity using the Van der Pauw equation: $\rho = (\pi t R/\ln 2)$, where $t$ is the estimated thickness of the sample~\cite{huang_2007_1}.  Given the uncertainty in the sample thickness, the estimated resistivity is accurate to within about a factor of two.  We have not attempted to account for the effect of changes in the sample thickness with pressure on the resistivity estimate.

The AC magnetic susceptibility measurements were performed using a balanced primary/secondary coil system~\cite{deemyad_2001_1}. The primary provides a root mean square field of $\sim 3$~gauss at 1000 Hz.  The coils are connected to two Stanford Research SRS830 lock-in amplifiers measuring at both the first and the third harmonic of the excitation frequency~\cite{Ishida1990}. The detection coil is connected through a Stanford Research SR554 transformer/preamplifier.  The third harmonic measurement is useful because it is less affected by background changes unrelated to the superconducting transition than the first harmonic signal.  Several different resistivity or ACS experiments were performed while taking different paths through the pressure-temperature phase space.  Table~\ref{table:run_summary} summarizes the different experiments that were performed.

\section{Results}
A phase diagram summarizing our results is presented in Figure~\ref{fig:phase}b.  Runs A and B were aimed primarily at clarifying the superconducting properties of Ba-V.  In these experiments pressure was increased at room temperature to pressures above 45~GPa, where the Ba(IV$\rightarrow$V) transition occurs.  For both set of measurements, during loading, the sample was cooled to low temperature several times and $T_c$ was measured. Once the maximum pressure was reached, the sample was annealed at room temperature for at least 6 hours before cooling in order to help ensure complete transformation of the sample to Ba-V.  The sample was then cooled to low temperature and maintained below $\sim 10$~K while subsequent measurements of $T_c$ were taken during unloading.

\begin{figure}	
	\includegraphics[width=\columnwidth]{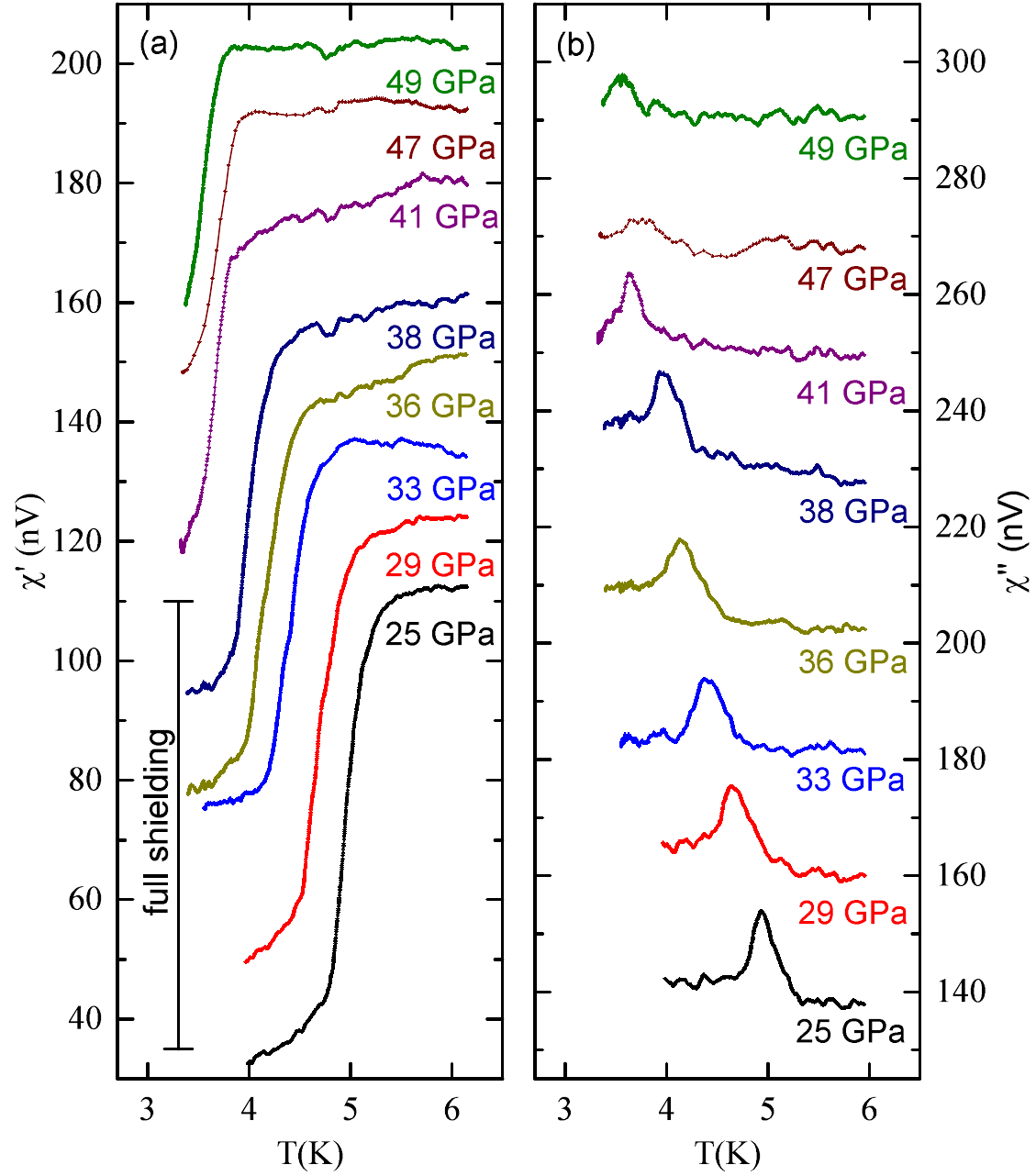}
    \caption{Real (a) and imaginary (b) part of the magnetic susceptibility for Run A while unloading  the pressure below 10 K. Each curve has been offset by a constant value for clarity. The size of the transitions are consistent with the estimated jump for a full shielding indicated by the scale bar at the bottom left of the figure.}
\label{fig:acs}
\end{figure}
Figure~\ref{fig:acs} presents a representative sample of our magnetic susceptibility data from Run A.  The data shown correspond to measurements taken while unloading pressure at low temperature.  Superconducting transitions are clearly visible both as drops in the real part of the susceptibility ($\chi'$) and peaks in the imaginary part of the susceptibility ($\chi''$).  From the geometry of the coil system, the approximate size and shape of the sample, and the calculated demagnetization factor for the sample~\cite{aharoni_1997_1}, we arrive at an estimate that full shielding should give a transition of about 75~nV in $\chi'$.  Our susceptibility data are thus consistent with bulk superconductivity.  The $T_c$ values plotted in Fig.~\ref{fig:phase}b have been taken from the midpoint of the $\chi'$ transition, and the error bar represents the 20\%-80\% transition width.

The susceptibility data from Runs A and B shows that $T_c$ varies smoothly with pressure across the Ba(IV$\rightarrow$V) transition with no discernible change in the slope $dT_c/dP$. In Run A, we reduced the pressure to 25~GPa at low temperature, which is lower that the critical pressure for the Ba(V$\rightarrow$VI) transition (30~GPa), yet we found that $T_c$ vs pressure agreed closely with the data obtained during room temperature loading~\footnote{The first data point obtained on loading for Run A ($P \sim 14$~GPa, $T_c \sim 1.8$~K) should correspond to the Ba-VI structure.  For this particular measurement, a pressure of 9~GPa was applied at room temperature.  During cooling the pressure increased to 14~GPa.  The pressure exceeded 12.5~GPa at at about 70~K.  This path will produce the metastable Ba-VI structure rather than Ba-IV.}.  The similarity of the room temperature loading and low temperature unloading curves across the Ba(V$\rightarrow$VI) transition suggests that either 1.\ Ba-IV, Ba-V, and Ba-VI all have nearly the same $T_c$ in the vicinity of 30~GPa, or 2.\ on unloading at low temperature the Ba(V$\rightarrow$VI) transition does not occur immediately so that the sample remains in the Ba-VI structure to pressures below 30~GPa.  We note that Desgreniers~\textit{et al.}~\cite{Desgreniers2015} did not seem to explore whether there is any hysteresis in the Ba(V$\rightarrow$VI) transition.  Below, we will see that Run C suggests that the latter of the two above possibilities is probably correct.

\begin{figure}
	\includegraphics[width=\columnwidth]{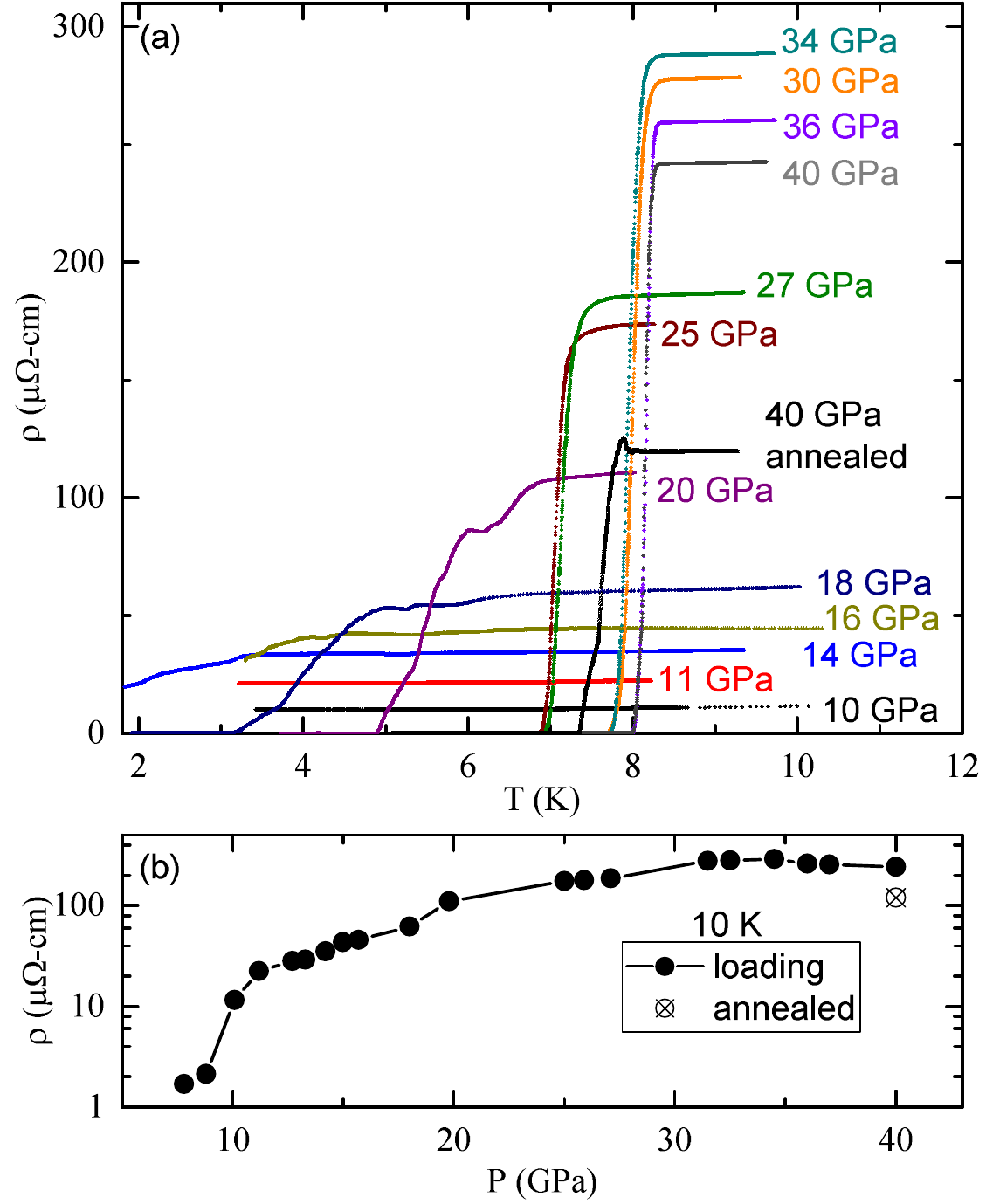}
    \caption{(a) Run C resistivity data taken while increasing the pressure at 10 K so as to enter the Ba-VI state. Superconductivity was observed beginning at 14 GPa with an incomplete transition above 2 K. The critical temperature peaked slightly above 8~K at 30~GPa. At 40~GPa the sample was annealed for 3 hours at 50 K, resulting in a slight drop in $T_c$ and broadening of the transition. (b) Resistivity at 10~K versus pressure for Run C.  The resistivity dropped by factor of $\sim 2$ after annealing at 50~K.  The changes in the magnitude of the resistivity and superconducting transition after annealing suggest that the sample has partially converted to Ba-V.}
\label{fig:rho}
\end{figure}

In Run C, we measured electrical resistivity and approached the Ba-VI structure from the low pressure side.  An initial pressure of $\sim 8$~GPa was applied at room temperature and then the sample was cooled to low temperature.  At this point the sample is expected to be the Ba-II structure.  Subsequent pressure increases were carried out while the temperature was kept below 10~K, which should drive the sample across the Ba(II$\rightarrow$VI) transition. Figure~\ref{fig:rho}a shows the electrical resistivity versus temperature for several different pressures.  At 10 and 11~GPa we observed no trace of superconductivity down to $\sim 3$~K.  Broad, incomplete superconducting transitions first appear at 14 GPa and the transitions become complete (reaching effectively zero resistance) for pressures at and above 18 GPa.  The transition remains broad to 20~GPa and then suddenly becomes much sharper at higher pressures.  Remarkably, we find that $T_c$ increases to $\sim 8$~K by 30 GPa, which is about 60\% higher than the highest $T_c$ observed in previous measurements in which pressure was applied at room temperature.

Above 30~GPa, which is the critical pressure for the Ba(VI$\rightarrow$V) transition, $T_c$ saturates at 8~K and remains roughly constant up to 40~GPa.  The fact that $T_c$ does not immediately drop to the values expected for Ba-V (4-5~K) at pressures above 30~GPa suggest that the Ba(VI$\rightarrow$V) transition is sluggish at these low temperatures.  We note that since $T_c$ was measured via resistivity, it is possible that part of the sample has converted to Ba-V, while enough Ba-VI remains to allow a super-current to percolate through the sample. This scenario is supported by the lower panel of Fig.~\ref{fig:rho}, which shows the resistivity versus pressure measured at 10~K.  The resistivity increases sharply in with pressure in Ba-VI and then saturates begins to decline above 30~GPa.  At 40~GPa, we raised the temperature to 50~K and annealed the sample for three hours before cooling back down to re-measure.  After this annealing process, the resistivity dropped sharply by a factor of two (crossed circle in Fig.~\ref{fig:rho}b), and the superconducting transition dropped by about half a kelvin and became substantially broader.  These observations support the conclusion that the pressure-driven Ba(VI$\rightarrow$V) transition is not immediate when the sample is maintained at low temperatures. 

\section{Discussion}
Using first principles calculations, Desgreniers~\textit{et al}.\ estimated $T_c = 3.6$~K at 16.2~GPa for the Ba-VI structure.  This value matches our results quite well - we found a resistive onset temperature of about 3.5~K at 16.5~GPa.  It does not seem that these calculations have been carried out at different pressures within the Ba-VI structure and it would be interesting to see if theory reproduces the positive $dT_c/dP$ trend that we have observed here.  Another theoretical effort addressed the Ba-V structure and computed values of $T_c$ ranging from 7.8~K to 9.5~K at 45~GPa~\cite{Da-Wei2013}. This is somewhat higher than the $T_c \sim 3.5$~K that we have observed at this pressure, however, the slope $dT_c/dP$ they compute ($\sim 0.1$~K/GPa) is roughly equal to what we find for the Ba-V in the pressure range 30-65~GPa.  The complexity of the Ba-IV crystal structure apparently makes first principles superconducting parameter computation intractable without the use of structural approximations.

Trends in the superconducting and structural behavior of the alkaline earths, as well as many other elements, can be understood in terms of pressure-induced transfer of $sp$-electrons into $d$-states~\cite{Hamlin2015}.  At ambient pressure, Be is marginally superconducting with a $T_c$ of only 26~mK, and none of the other alkaline earths are superconducting.  Under high pressure, the light alkaline earth elements (Be and Mg) do not show any enhancement of superconductivity, while the heavy alkaline earths (Ca, Sr, and Ba) all become superconducting.  This difference in behavior derives from the fact that Be and Mg have no nearby $d$-states, while Ca, Sr and Ba do.  As pressure increases, the $s$-electrons in Ca, Sr, and Ba transfer into $d$-bands so that these elements become transition-metal-like at high pressure.  The substantial density of states at the Fermi-level due to these $d$ states is favorable for superconductivity.  Across the heavy alkaline earths, superconductivity only seems to appear when the number of $d$ electrons, $N_d$, exceeds $\sim 0.92$~\cite{Skriver1982}.  In addition, the appearance of complex modulated structures in the alkaline earths has been connected with the near completion $s\rightarrow d$ transfer~\cite{Loa2012}.  The completion of $s\rightarrow d$ transfer, the onset of superconductivity, and the appearance of modulated structures all occur at progressively higher pressures going from Ba, to Sr, to Ca~\cite{Hamlin2015}.  Barium at lower pressures can thus be considered as a close analogue of Sr and Ca at higher pressures.

The present results show that a significant $T_c$ enhancement can be obtained by avoiding the complex modulated Ba-IV structure.  Why might this be, and what does this imply for the superconducting phase diagrams of Ca, Sr, and other elements exhibiting pressure-induced complex modulated crystal structures?  The modulated structures that appear in several elements have been attributed to the formation of a charge density wave (CDW)~\cite{degtyareva_2007_1,Loa2009,Nagara2010}.  A synonymous interpretation views these structures as deriving from Fermi-surface/Brillouin-zone interactions that lead to the opening of pseudo-gap at the Fermi level~\cite{Degtyareva2003,Ackland2004,Degtyareva2006}.  Both CDW/pseudo-gap states and superconductivity rely on a modification of the density of states at the Fermi-level, and this can result in a competition between the two phases such that suppressing the CDW/pseudo-gap leads to an enhancement of $T_c$~\cite{Bilbro1976}.  Among the elemental solids, S is one of the prototypes for this behavior: The increase of $T_c$ in the complex S-IV structure has been related the suppression and then destruction of a CDW~\cite{degtyareva_2007_1,Monni2017}.  In the case of U, the suppression of several CDW transitions correlates with a maximum in the $T_c$ vs pressure curve~\cite{Maple1972,Lander1994,McMahon2006}.  For P, calculations involving a commensurate approximation of the incommensurate P-IV structure indicate a suppression of $T_c$ in the modulated structure compared to nearby simple cubic and simple hexagonal structures~\cite{Nakanishi2012}.  The $T_c$ suppression was attributed to a lowering of the density of states due to the modulation.  A similar scenario may be at play for Ba such that the $T_c$ of Ba-IV is suppressed due to the presence of a pseudo-gap.  On the other hand, Loa~\textit{et al.}~\cite{Loa2012} performed electronic structure calculations for a simplified approximation of the Ba-IVc structure and found no indication of a pseudo-gap.  Furthermore, it is possible that the higher $T_c$ of Ba-VI relative to Ba-IV derives from gross differences in the crystal structures, rather than the presence of a pseudo-gap in Ba-IV.  Low temperature compression (and decompression) experiments in elements including P~\cite{Kawamura1985} and Si~\cite{Valyanskaya1993} have revealed $T_c$ enhancements that are likely unrelated to the type of CDW/superconductivity competition described above.

Nonetheless, the present results and the above factors hint that other elements crystallizing in complex modulated structures at room temperature might exhibit enhanced superconducting critical temperatures when compressed at low temperature.  Desgreniers~\textit{et al}.~\cite{Desgreniers2015} proposed that systems exhibiting complex modulated structures structures at ambient temperatures may show a general tenancy to adopt alternative metastable structures when pressurized cryogenically.  The kinetics of solid-solid phase transitions are still incompletely understood~\cite{Pogatscher2016} so that it is an interesting and open question whether or not such metastable structures will appear in any particular case.  Further experiments probing the effects of low temperature compression on both the structural and superconducting properties of elements with incommensurate/modulated structures may offer a new perspective on the occurrence of superconductivity throughout the periodic table.  Given the closely similar superconducting/structural phase diagrams of the heavy alkaline earths, low temperature compression experiments are particularly well motivated for Ca, which exhibits the highest critical temperature among the elements ($T_c > 21$~K) in a host-guest structure at pressures above 210~GPa~\cite{Sakata2011}.

\section*{Acknowledgments}
Development of \textit{in situ} pressure tuning equipment partially was supported by The National High Magnetic Field Laboratory User Collaboration Grants Program.  The National High Magnetic Field Laboratory is supported by National Science Foundation Cooperative Agreement No. DMR-1157490 and the State of Florida.  Measurements were supported by NSF grant DMR-1453752.  Designer diamond anvils supported by DOE-NNSA Grant No.\ DE-NA0002928 and under the auspices of the U.S. Department of Energy by Lawrence Livermore National Laboratory under Contract DE-AC52-07NA27344. We thank A.\ Linscheid for helpful conversations.

\bibliography{Ba_references}
\end{document}